# Sampling Techniques in Bayesian Finite Element Model Updating

I. Boulkaibet [a], T. Marwala[a], L. Mthembu[a], M. I. Friswell[b], S. Adhikari [c]

[a] The Centre For Intelligent System Modelling (CIMS), University of Johannesburg, SA.
[b] Aerospace Structures, College of Engineering, Swansea University, UK.
[c] Aerospace Engineering, College of Engineering, Swansea University, UK.

**Abstract**

Recent papers in the field of Finite Element Model (FEM) updating have highlighted the benefits of Bayesian techniques. The Bayesian approaches are designed to deal with the uncertainties associated with complex systems, which is the main problem in the development and updating of FEMs. This paper highlights the complexities and challenges of implementing any Bayesian method when the analysis involves a complicated structural dynamic model. In such systems an analytical Bayesian formulation might not be available in an analytic form; therefore this leads to the use of numerical methods, i.e. sampling methods. The main challenge then is to determine an efficient sampling of the model parameter space. In this paper, three sampling techniques, the Metropolis-Hastings (MH) algorithm, Slice Sampling and the Hybrid Monte Carlo (HMC) technique, are tested by updating a structural beam model. The efficiency and limitations of each technique is investigated when the FEM updating problem is implemented using the Bayesian Approach. Both MH and HMC techniques are found to perform better than the Slice sampling when Young's modulus is chosen as the updating parameter. The HMC method gives better results than MH and Slice sampling techniques, when the area moment of inertias and section areas are updated.

**1. Introduction**

Finite element models (FEM) use a numerical method to provide approximate solutions for complex engineering problems [1, 2]. The FEM method is recognised as a powerful method for computing displacements, stresses and strains in structures under a set of loads. However, accurate solutions obtained by an FEM are only possible for few simple model cases. In complex systems FEM results are different from those obtained from experiments [3, 4]. These differences can be a result of the modelling errors and/or the uncertainties associated with the process of constructing the FEM of the structure. Therefore, the models obtained from the finite element method need to be updated to match the measurements. In the recent years, Bayesian model updating techniques have shown promising results in systems identification research [4, 6, 7].

The Bayesian approach provides accurate solutions for complex systems, due to its ability to characterize the uncertainties of these complex systems. In this paper, the Bayesian approach is used to update the uncertain parameter of the FEM where these can be represented as random vectors with a joint probability density function (pdf). This posterior (pdf) characterises the uncertainty in the model parameters. The use of Bayesian techniques becomes useful when an analytical solution to this function is not available. This is often the case because of the high dimensionality of the parameter search space. In these situations some Markov chain Monte Carlo (MCMC) simulation methods provide the only practical solution [4, 7].

In this paper three sampling techniques, the Metropolis-Hastings (MH), the Slice Sampling (SS) and the Hybrid Monte Carlo (HMC) are investigated on their ability to sample the posterior pdf of FEM updating parameters. These three techniques are tested by updating a structural beam model. The efficiency, reliability and limitations of each technique are investigated when a Bayesian approach is implemented on a FEM updating problem.

In the next section, the finite element model background is presented. In Section 3, an introduction to the Bayesian framework is introduced where the posterior distribution of the uncertain parameters of FEM is also presented. Section 4 introduces the sampling techniques used to predict the posterior distribution. Section 5 presents an implementation of the

Bayesian FEM updating on a simple cantilever steel beam, where two cases of study will be provided. Section 6 concludes the paper.

**2 Finite Element Model Background**

In finite element modelling, an *N* degree of freedom dynamic structure may be described by the matrix equation of motion [14]

$$\mathbf{M}\ddot{x}(t) + \mathbf{C}\dot{x}(t) + \mathbf{K}x(t) = \mathbf{f}(t) \tag{1}$$

where $\mathbf{M}, \mathbf{C}$ and $\mathbf{K}$ are the mass, damping and stiffness matrices of size $N \times N$. $\mathbf{x}(t)$ is the vector of *N* degrees of freedom and $\mathbf{f}(t)$ is the vector of forces applied to the structure. In case that no external forces are applied to the structure and if the damping terms are neglected ($\mathbf{C} \sim \mathbf{0}$), the dynamic equation may be written in the modal domain (natural frequencies and mode shapes) where the error vector for the *i*th mode is obtained from

$$[-(\omega_i^m)^2 \mathbf{M} + \mathbf{K}]\boldsymbol{\phi}_i^m = \boldsymbol{\varepsilon}_i \tag{2}$$

$\omega_i^m$ is the *ith* measured natural frequency, $\boldsymbol{\phi}_i^m$ is the *ith* measured mode shape vector and $\boldsymbol{\varepsilon}_i$ is the *ith* error vector. In Equation (2), the error vector $\boldsymbol{\varepsilon}_i$ is equal to $\mathbf{0}$ if the system matrices $\mathbf{M}$ and $\mathbf{K}$ correspond to the modal properties ($\omega_i^m$ and $\boldsymbol{\phi}_i^m$). However, $\boldsymbol{\varepsilon}_i$ is a non-zero vector if the system matrices obtained analytically from the finite element model do not match the measured modal properties $\omega_i^m$ and $\boldsymbol{\phi}_i^m$. In the case that Equation (2) contains both measured vectors, $\omega_i^m$ and $\boldsymbol{\phi}_i^m$, and the matrices obtained from the analytic methods $\mathbf{M}$ and $\mathbf{K}$, the error obtained in Equation (2) represents the differences between the measured and the analytic modal properties. Another problem might arise in many practical situations, where the dimension of the mode shape vector does not match the dimension of the system matrices. This is because the measured modal coordinates are fewer than the finite element modal coordinates. To ensure the compatibility between the system matrices and the mode shape vectors, the dimension of the system matrices is reduced using Guyan reduction [15].

**3 Bayesian Inferences**

In order to update the mathematical models, the uncertain parameters have to be identified. The Bayes theorem offers this possibility, where the uncertain parameters can be determined from their measurements [4, 5]. In this work, the Bayesian method is used to solve the finite element updating problem based on modal properties. Bayesian approaches are governed by Bayes rule [5, 8]:

$$P(\mathbf{E}|D) = \frac{P(D|\mathbf{E})P(E)}{P(D)} \tag{2}$$

where $\mathbf{E}$ represent the vector of updating parameters and the mass $\mathbf{M}$ and stiffness $\mathbf{K}$ matrices are functions of the updating parameters $\mathbf{E}$. The quantity $P(\mathbf{E})$, known as the prior probability distribution, is a function of the updating parameters in the absence of the data. $D$ is the measured modal properties; the natural frequencies $\omega_i$ and mode shapes $\boldsymbol{\phi}_i$. The quantity $P(\mathbf{E}|D)$ is the posterior probability distribution function of the parameters in the presence of the data $D$. $P(D|\mathbf{E})$ is the likelihood probability distribution function and $P(D)$ is a normalization factor [4,5,8].

**3.1 The Likelihood Distribution Function**

The likelihood distribution can be seen as the probability of the modal measurements in the presence of uncertain parameters. The likelihood function $P(D|\mathbf{E})$ can be defined as the normalized exponent of the error function that represents the differences between the measured and the analytic frequencies.

This function can be written as follows

$$P(D|\mathbf{E}) = \frac{1}{Z_D(\beta)} exp\left(-\beta \sum_j^F \sum_i^{N_m} (\varepsilon_{ij}^2)\right) \tag{4}$$

where, $\beta$ is the coefficient of the measured modal data contribution to the error. $\varepsilon_{ij}$ represents the error between the measured and analytical frequencies where *i* indicates the *i*th modal properties and subscript *j* represents the *j*th measurement position. The superscript *F* is the number of measured mode shape coordinates; $N_m$ is the number of measured modes and $Z_D(\beta)$ is a normalising constant given by

$$Z_D(\beta) = \int exp\left(-\beta \sum_j^F \sum_i^{N_m}(\varepsilon_{ij}^2)\right) d[D] = \left(\frac{\pi}{\beta}\right)^{FN_m/2} \tag{5}$$

### 3.2 Prior Distribution Function

The prior pdf represents the prior knowledge of the updating parameters **E**. It quantifies the uncertainty of the parameters **E**. In this paper, some parameters are updated more intensely than others. For example, parameters next to joints should be updated more intensely than for those corresponding to smooth surface areas far from joints. The prior probability distribution function for parameters **E** is assumed to be Gaussian and is given by [8, 9]

$$P(\mathbf{E}) = \frac{1}{Z_{\mathbf{E}}(\alpha)} exp\left(-\sum_i^Q \frac{\alpha_i}{2} \|\mathbf{E}\|^2\right) \tag{6}$$

where $Q$ is the number of groups of parameters to be updated, and $\alpha_i$ is the coefficient of the prior pdf for the $i$th group of updating parameters. The $\|*\|$ is the Euclidean norm of $*$. In Equation (6), if $\alpha_i$ is constant for all of the updating parameters, then the updated parameters will be of the same order of magnitude. Equation (6) may be viewed as a regularization parameter [10]. In Equation (6), Gaussian priors were conveniently chosen because many natural processes tend to have a Gaussian distribution. The function $Z_{\mathbf{E}}(\alpha)$ is a normalization factor given by [4, 8]

$$Z_{\mathbf{E}}(\alpha) = \int exp\left(-\sum_i^Q \frac{\alpha_i}{2} \|\mathbf{E}\|^2\right) d[\mathbf{E}] = (2\pi)^{Q/2} \prod_{i=1}^Q \frac{1}{\sqrt{\alpha_i}} \tag{7}$$

### 3.3 Posterior Probability Distribution Function

The posterior distribution function of the parameters **E** given the observed data $D$ is denoted as $P(\mathbf{E}|D)$ and obtained by applying the Bayes' theorem as represented in Equation (3). The distribution $P(\mathbf{E}|D)$ is calculated by substituting Equations (4) and (6) into Equation (3) to give [4]

$$P(\mathbf{E}|D) = \frac{1}{Z_s(\alpha,\beta)} exp\left(-\beta \sum_j^F \sum_i^{N_m}(\varepsilon_{ij}^2) - \sum_i^Q \frac{\alpha_i}{2} \|\mathbf{E}\|^2\right) \tag{8}$$

where

$$Z_s(\alpha,\beta) = Z_D(\beta).Z_{\mathbf{E}}(\alpha) \tag{9}$$

### 4. Sampling Techniques

Sampling techniques are very useful numerical methods which can be employed to simplify the Bayesian inference by providing a set of random samples from posterior distribution [5, 7, 8, 11]. Suppose that $Y$ is the observation of certain parameters at different discrete time instants; the probabilistic information for the prediction of the future responses $Y$ at different time instants is contained in the robust predictive PDF which is given by the Theorem of Total Probability as

$$P(Y|D) = \int P(Y|\mathbf{E})P(\mathbf{E}|D)d[\mathbf{E}] \tag{10}$$

Equation (10) depends on the posterior distribution function which is very difficult to solve analytically due to the dimension of the updating parameters. Therefore, sampling techniques, such as Markov chain Monte Carlo (MCMC) methods, are employed to predict the updating parameter distribution and subsequently to predict the modal properties. Given a set of $N_s$ random parameter vector drawn from a pdf $P(\mathbf{E}|D)$, the expectation value of any observed function $Y$ can be easily estimated.

The integral in Equation (10) is solved using three different sampling techniques: the MH algorithm, the slice sampling algorithm and the HMC algorithm [5, 7, 12, 13]. These algorithms are used to generate a sequence of vectors $\{\mathbf{E}_1, \mathbf{E}_2, \dots, \mathbf{E}_{N_s}\}$ where $N_s$ is the number of samples and these vectors can be used to form a Markov chain. This generated vector is then used to predict the form of the stationary distribution function $P(\mathbf{E}|D)$. The integral in Equation (10) can be approximated as

$$\tilde{Y} \cong \frac{1}{N_s} \sum_{i=1}^{N_s} G(\mathbf{E}_i) \tag{11}$$

where $G$ is a function that depends on the updated parameters $\mathbf{E}_i$. As an example, if $\mathbf{G} = \mathbf{E}$ then $\tilde{Y}$ becomes the expected value of **E**. Generally, $\tilde{Y}$ is the vector that contains the modal properties and $N_s$ is the number of retained states.

*4.1 The Metropolis-Hastings Algorithm*

The MH algorithm is one of the simplest sampling methods, and is related to rejection and importance sampling [5]. To sample from the posterior distribution function $P(\mathbf{E}|D)$, where $\mathbf{E} = \{E_1, E_2, \ldots, E_{N_m}\}$ is a $N_m$-dimensional parameter vector, a proposal density distribution $q(\mathbf{E}|\mathbf{E}_{t-1})$ is introduced in order to generate a random vector $\mathbf{E}$ given the value at the previous iteration of the algorithm. The MH algorithm consists of two basic stages: the draw from the proposed density stage and the retained/rejected stage. The MH algorithm can be summarized as:

1) An initial value $\mathbf{E}_0$ is used to initiate the algorithm.
2) At iteration $t$, $\mathbf{E}^*$ is drawn from the proposed density $q(\mathbf{E}|\mathbf{E}_{t-1})$, where $\mathbf{E}_{t-1}$ is the parameter value at the previous step.
3) Update the FEM to obtain the new analytic frequencies, then compute the acceptance probability, given by

$$a(\mathbf{E}^*, \mathbf{E}_{t-1}) = \min\left\{1, \frac{P(\mathbf{E}^*|D)q(\mathbf{E}_{t-1}|\mathbf{E}^*)}{P(\mathbf{E}_{t-1}|D)q(\mathbf{E}^*|\mathbf{E}_{t-1})}\right\}$$

4) Draw $u$ from the uniform distribution $U(0,1)$.
5) If $u \leq a(\mathbf{E}^*, \mathbf{E}_{t-1})$ accept $\mathbf{E}^*$. Otherwise, reject $\mathbf{E}^*$
6) Return to step (2)

**4.2 The Slice Sampling Algorithm**

The slice sampling method is a simple technique that can provide an adaptive step size, which is automatically adjusted to match the characteristics of the posterior distribution function [5, 12]. In this method, the goal is to sample uniformly from the area under the posterior distribution $P(\mathbf{E}|D)$ where $\mathbf{E} = \{E_1, E_2, \ldots, E_{N_m}\}$. The algorithm of this technique can be described as [12]:

1) Draw $Y$ from the uniform distribution $U(0, P(\mathbf{E}_0|D))$.
2) Initiate the interval around $\mathbf{E}_0$ as follows:

   For $i = 1$ to $N$
   $U_i \sim Uniform(0,1)$
   $L_i \leftarrow E_{0,i} - w_i U_i$
   $R_i \leftarrow L_i + w_i$
   End

3) Sample from the interval $I = (R, L)$ and do the following:
   Repeat:
   For $i = 1$ to $N$
   $U_i \sim Uniform(0,1)$
   $E_{1,i} \leftarrow L_i + U_i(R_i - L_i)$
   End
   IF $Y \leq P(\mathbf{E}_0|\mathbf{E})$ Then exit loop
   For $i = 1$ to $N$
   IF $E_{1,i} < E_{o,i}$ Then $L_i \leftarrow E_{1,i}$ Else $R_i \leftarrow E_{1,i}$

4) Repeat step (3) to get $N_s$ samples.

**4.3 Hybrid Monte Carlo**

The Hybrid Monte Carlo method, known as the Hamiltonian Markov Chain method, is a good method for solving higher-dimensional complex problems [5, 7, 13]. In HMC, a new dynamical system is considered in which auxiliary variables, called momentum, $\mathbf{p} \in R^N$, are introduced and the uncertain parameters, $\mathbf{E} \in R^N$, in the target posterior distribution function are treated as displacements. The total energy, Hamiltonian function, of the new dynamical system is defined by $H(\mathbf{E}, \mathbf{p}) = V(\mathbf{E}) + W(\mathbf{p})$, where the potential energy is defined by $V(\mathbf{E}) = -\ln(P(\mathbf{E}|D))$ and the kinetic energy $W(\mathbf{p}) = \mathbf{p}^T \mathbf{M}^{-1} \mathbf{p}/2$ depends only on $\mathbf{p}$ and some chosen positive definite matrix $\mathbf{M} \in R^{N \times N}$. Using Hamilton's equations, the evolution of $(E, p)$ through time $t$ and time step $\delta t$ is given by

$$\mathbf{p}\left(t + \frac{\delta t}{2}\right) = \mathbf{p}(t) - \frac{\delta t}{2}\nabla V[\mathbf{E}(t)] \qquad (12)$$

$$\mathbf{E}(t + \delta t) = \mathbf{E}(t) + \delta t \mathbf{M}^{-1} \mathbf{p}\left(t + \frac{\delta t}{2}\right) \qquad (13)$$

$$\mathbf{p}(t+\delta t) = \mathbf{p}\left(t+\frac{\delta t}{2}\right) - \frac{\delta t}{2}\nabla V[\mathbf{E}(t+\delta t)] \quad (14)$$

where $\nabla V$ is obtained numerically by finite difference as

$$\frac{\partial V}{\partial E_i} = \frac{V(\mathbf{E}+\Delta h)-V(\mathbf{E}-\Delta h)}{2h\Delta_i} \quad (15)$$

$\Delta = [\Delta_1, \Delta_2, \ldots, \Delta_N]$ is the perturbation vector, the distribution of which is user-specified and $h$ is a scalar which dictates the size of the perturbation of $\mathbf{E}$. After each iteration of Equations (12)-(14), the resulting candidate state is accepted or rejected according to the Metropolis criterion based on the value of the Hamiltonian $H(\mathbf{E}, \mathbf{p})$. Thus, if $(\mathbf{E}, \mathbf{p})$ is the initial state and $(\mathbf{E}^*, \mathbf{p}^*)$ is the state after the equations above have been updated, then this candidate state is accepted with probability $\min(1, exp\{H(\mathbf{E}, \mathbf{p}) - H(\mathbf{E}^*, \mathbf{p}^*)\})$. The new vector $\mathbf{E}$ will be used for the next iteration and the algorithm will be stopped when a $N_s$ samples of $\mathbf{E}$ are provided.

## 5 Beam Example

An experimental cantilever steel beam is updated based on the measurements of Kraaij [16]. The beam has the following dimensions: length 500 mm, width 60 mm and thickness 10 mm. $E = 2.1 \times 10^{11} \text{N/m}^2$, $v = 0.3$ and $\rho = 7850 \text{ kg/m}^3$. Three accelerometers were used in the experiment, which were all located 490 mm from the clamped end. This location is chosen because the response on this point is large [16]. Each accelerometer has a mass of 40g; the middle accelerometer is of type $303A3$, the outer accelerometers are of type $303A2$ (see [16] for more details of experimental set-up).

To test the updating methods, the beam was modeled using Version 6.3 of the Structural Dynamics Toolbox SDT® for MATLAB. The beam was divided into 50 Euler–Bernoulli beam elements and excited at various positions. The measured natural frequencies of interest of this structure are: 31.9 Hz, 197.9 Hz, 553 Hz, 1082.2 Hz and 1781.5Hz, which correspond to modes 1, 3, 5, 7, and 9, respectively. For the first set of experiments, the Young's modulus of the beam elements was used as an updating parameter where for every 10 elements a different Young's modulus is allocated. Thus, the parameters to be updated can be represented by a vector of 5 variables $\mathbf{E} = \{E_1, E_2, E_3, E_4, E_5\}$. In the second set of experiments, the moments of inertia and the section areas are updated. This is done by associating an area moment of inertia, $I_x$, and an area, $A_x$, to every 25 elements of the beam (this will reduce the number of the parameters to be updated to four parameters). The updated parameters vector is thus $\mathbf{E} = \{I_{x1}, I_{x2}, A_{x1}, A_{x2}\}$.

### 5.1 Updating Young's modulus

In this section, a vector of 5 parameters $\mathbf{E} = \{E_1, E_2, E_3, E_4, E_5\}$ is updated using the Bayesian approach. The reason for using a large number of updating parameters is to determine the performance and the convergence speed of each sampling technique when a large number of variables are introduced in the updating process. $N_s$ samples of the vector $\mathbf{E}$ were generated from the posterior distribution function, $P(\mathbf{E}|D)$ mentioned in Equation (8). The constant $\beta$ in Equation (8) was set equal to 1, and the coefficients $\alpha_i$ were set equal to $\frac{1}{\sigma_i^2}$, where $\sigma_i^2$ is the variance of the parameter $E_i$. Since the updating parameter vector contains only the Young's modulus, all $\sigma_i$ were set equal to $2 \times 10^{11}$ (a large value of $\sigma_i$ to weight the value of $E$). The updating parameters $E_i$ were bounded with a maximum equal to $2.5 \times 10^{11}$ and a minimum equal to $1.7 \times 10^{11}$. The number of samples $N_s$ was 1000, for all techniques, and the initial vector of $\mathbf{E}$ is $\{2.4 \times 10^{11}, 2.4 \times 10^{11}, 2.4 \times 10^{11}, 2.4 \times 10^{11}, 2.4 \times 10^{11}\}$. Instead of using the mean steel value of $\mathbf{E}$ as an initial value, a large value of the initial parameter vector is chosen to highlight the updating process. The results for the updating vector and the frequencies are given in Tables 1 and 2.

|  | Young's modulus (N/m²) | | | |
|---|---|---|---|---|
|  | Initial | M-H Method | Slice Sampling Method | HMC Method |
| $E_1$ | $2.4 \times 10^{11}$ | $2.177 \times 10^{11}$ | $2.387 \times 10^{11}$ | $1.727 \times 10^{11}$ |
| $E_2$ | $2.4 \times 10^{11}$ | $2.151 \times 10^{11}$ | $1.851 \times 10^{11}$ | $2.344 \times 10^{11}$ |
| $E_3$ | $2.4 \times 10^{11}$ | $1.944 \times 10^{11}$ | $1.849 \times 10^{11}$ | $1.838 \times 10^{11}$ |
| $E_4$ | $2.4 \times 10^{11}$ | $1.785 \times 10^{11}$ | $1.853 \times 10^{11}$ | $2.052 \times 10^{11}$ |
| $E_5$ | $2.4 \times 10^{11}$ | $2.036 \times 10^{11}$ | $1.864 \times 10^{11}$ | $2.004 \times 10^{11}$ |

Table 1: The updated vector of Young's modulus using MH, Slice Sampling and HMC techniques.

The MH technique updates all vector parameters simultaneously and gives results that are close to the mean value for steel of $2.1 \times 10^{11} \text{N/m}^2$. The same comment can be made for the HMC technique where three of these parameters are close to the mean value. The Slice sampling method gives updating parameters that are far from the mean, because of the way that the Slice technique generates samples. The Slice sampling technique sequentially updates individual vector entries as oppose to updating all entries simultaneously, see section 4. In this case, a small adjustment of the first parameter can cause a significant updating of the rest of the updating vector.

| Modes | Measured Frequency (Hz) | Initial Frequency (Hz) | Error (%) | Frequencies M-H Method (Hz) | Error (%) | Frequencies Slice Sampling Method (Hz) | Error (%) | Frequencies HMC Method (Hz) | Error (%) |
|---|---|---|---|---|---|---|---|---|---|
| 1 | 31.9 | 32.7 | 2.51 | 30.8 | 3.38 | 30.8 | 3.44 | 29 | 9 |
| 2 | 197.9 | 209.4 | 5.83 | 190.3 | 3.85 | 190.9 | 3.51 | 186.2 | 5.92 |
| 3 | 553 | 594.8 | 7.55 | 541.1 | 2.15 | 535.2 | 3.21 | 548 | 0.9 |
| 4 | 1082.2 | 1237.2 | 14.32 | 1123.6 | 3.83 | 1127.6 | 4.2 | 1099.99 | 1.64 |
| 5 | 1781.5 | 1961.7 | 10.12 | 1798.2 | 0.93 | 1772.4 | 0.51 | 1773.6 | 0.45 |

Table 2: Frequencies and Errors when MH, Slice Sampling and HMC techniques used to update Young's modulus.

The results in Table 2 show that the three sampling techniques give results that on average are better than the initial FEM. The three algorithms weight the measured frequencies in different ways because each algorithm has a different way of generating samples. In the Slice sampling technique, each variable is changed one at a time which is not the case with the M-H algorithm where all the parameters are varied at once. The HMC technique uses additional parameters to evaluate the sampling.

The MH and HMC algorithms give better results than those obtained by using the Slice Sampling method. The error between the third measured natural frequency and that of the initial model was 7.55%. When the MH method was used, this error was reduced to 2.15% and in using the Slice sampling method it was only reduced to 3.21%. The same comment can be made for the first and fourth natural frequencies. The errors for the second and fifth modes are targeted by the Slice sampling method, where the error is a little smaller than that obtained from the MH method. However,, the overall result shows that the MH method gave better results than the Slice sampling method in terms of errors. Also the convergence of the MH method was faster than the Slice sampling method. The results show that Slice sampling is inefficient in sampling Young's modulus when 1000 samples are generated. The HMC method converges faster than the other methods (MH and Slice sampling techniques). In addition, the HMC method gives a bad error for the two first modes where the error between the first measured natural frequency and that of the initial model was 2.51%. When the HMC method was used, this error was increased to 9% (similarly for the second mode). However, for the other three modes the HMC method gave better results than both the MH and Slice sampling methods. For example, the error between the third measured natural frequency and that of the initial model was 7.55%; the HMC method reduced this error to 0.9% whereas the Slice sampling method only reduced it to 3.21%. The same comment can be made in using MH algorithm where the error for the third mode only reduced to 2.15%. The three methods did not improve the first natural frequency because the same coefficient, $\beta$, was set for all of the natural frequencies. Choosing $\beta$ as a vector, to weight the natural frequencies, would improve the first natural frequency results. In general, the error between the measured frequency and these obtained by the three algorithms is a bit high when the Young's modulus is updated. In this case, conclusions about the most efficient method are difficult. For this reason, another updating exercise is performed, where the area moments of inertia and cross-section area are updated and the results obtained are discussed in Section 5.2.

**5.2 Updating Area Moments of Inertia and Cross-section Area**

In this section, four parameters $\mathbf{E} = \{I_{x1}, I_{x2}, A_{x1}, A_{x2}\}$ are updated using the Bayesian approach. $N_s$ samples of the vector $\mathbf{E}$ were generated from the posterior distribution function $P(\mathbf{E}|D)$. The constant $\beta$ is equal 1, and all coefficients $\alpha_i$ are set equal to $\frac{1}{\sigma_i^2}$, where $\sigma_i^2$ is the variance of the parameter $E_i$. Since the updating parameter vector contains area moments of inertia and section areas, the vector of $\sigma_i$ is defined as $\boldsymbol{\sigma} = [5 \times 10^{-9}, 5 \times 10^{-9}, 5 \times 10^{-4}, 5 \times 10^{-4}]$. The Young's modulus is set to $2.1 \times 10^{11} \text{N/m}^2$ as opposed to $2.4 \times 10^{11} \text{ N/m}^2$ in the previous section. The updating parameters $E_i$ are bounded by maximum values equal to $[7.5 \times 10^{-9}, 7.5 \times 10^{-9}, 9 \times 10^{-4}, 9 \times 10^{-4}]$ and minimum values equal to $[3.5 \times 10^{-9}, 3.5 \times 10^{-9}, 4.5 \times 10^{-4}, 4.5 \times 10^{-4}]$. The number of samples $N_s$ is set to 1000. The results are given in Tables 3 and 4.

|          | Initial **E**          | **E** vector, M-H Method | **E** vector, Slice Sampling Method | **E** vector, HMC Method |
|----------|------------------------|--------------------------|-------------------------------------|--------------------------|
| $I_{x1}$ | $5 \times 10^{-9}$     | $7.05 \times 10^{-9}$    | $7.88 \times 10^{-9}$               | $6.96 \times 10^{-9}$    |
| $I_{x2}$ | $5 \times 10^{-9}$     | $6.36 \times 10^{-9}$    | $3.67 \times 10^{-9}$               | $6.41 \times 10^{-9}$    |
| $A_{x1}$ | $8 \times 10^{-4}$     | $8.28 \times 10^{-4}$    | $5.62 \times 10^{-4}$               | $8.92 \times 10^{-4}$    |
| $A_{x2}$ | $8 \times 10^{-4}$     | $8.46 \times 10^{-4}$    | $8.09 \times 10^{-4}$               | $8.13 \times 10^{-4}$    |

Table 3: The updated vector using MH, Slice Sampling and HMC techniques.

The MH and HMC techniques update all parameters simultaneously and again give results that are closer to the the initial parameter vector than those obtained by using the Slice sampling technique (see Table 3). The Slice sampling method gives updating parameters far from the mean value of **E** and this because of the way that the Slice sampling technique updates (the Slice sampling technique updates each parameter in turn).

| Modes | Measured Frequency (Hz) | Initial Frequency (Hz) | Error (%) | Frequencies M-H Method (Hz) | Error (%) | Frequencies Slice Sampling Method (Hz) | Error (%) | Frequencies HMC Method (Hz) | Error (%) |
|-------|-------------------------|------------------------|-----------|-----------------------------|-----------|----------------------------------------|-----------|-----------------------------|-----------|
| 1 | 31.9   | 32     | 0.32 | 31.3   | 1.97 | 32.9   | 3.14 | 31.6   | 1.08 |
| 2 | 197.9  | 203.5  | 2.58 | 195.2  | 1.34 | 183.3  | 7.1  | 193.8  | 2.06 |
| 3 | 553    | 575.7  | 4.1  | 552    | 0.19 | 556.4  | 0.62 | 545.3  | 1.4  |
| 4 | 1082.2 | 1136.8 | 5.04 | 1087.6 | 0.5  | 1084   | 0.16 | 1077   | 0.48 |
| 5 | 1781.5 | 1889.6 | 6.06 | 1811.2 | 1.67 | 1799.4 | 1.00 | 1790.7 | 0.51 |

Table 4: Frequencies and Errors when MH, Slice Sampling and HMC techniques used to update momentums and area sections.

In this section, the results obtained give lower errors than those in Table 2. The results in Table 4 show that the HMC and HM methods give better updating than the Slice sampling algorithm. This can be seen by comparing the errors for modes 1 and 2. The error between the second measured natural frequency and that of the initial model was 2.58%. When the MH and HMC methods were used, the error was reduced to 1.34% and 2.06% respectively, and in using the Slice sampling method it was increased to 7.1%. Moreover, the results in Table 2 give a priority to HMC over MH and Slice sampling algorithms in term of convergence time since the HMC approach converges faster than the other methods. In addition, the HMC technique gives a small error compared to that obtained by using MH algorithm, where in modes 1, 4 and 5 the HMC method gives small frequency errors compared to the MH method. Despite the errors that may occur when the gradients are evaluated numerically, the HMC algorithm shows good results.

## 6. Conclusion

In this paper the finite element model updating problem is posed as a Bayesian problem. This means the uncertainty associated with the model parameters and the distribution of the data with these parameters is concisely formulated in Bayes theorem. To evaluate the resultant high dimensional posterior distribution three MCMC sampling techniques are implemented; the Metropolis-Hastings algorithm, Slice Sampling and the Hybrid Monte Carlo technique.

These sampling techniques are tested on a simple beam structure. In the first simulation the Hybrid Monte Carlo and Metropolis-Hastings technique gave more accurate results than the Slice sampling method. In addition the HMC method converges faster than both the MH and Slice sampling algorithms. In the second simulation the HMC method gave better results than both the MH and slice sampling methods and has more attractive convergence properties than the latter. Further work includes testing these algorithms on more complicated dynamic structures.